\documentclass[preprint, superscriptaddress, showpacs, longbibliography]{revtex4-1}
\usepackage{color, graphicx}
\usepackage{amssymb,amsmath, amsfonts, bm}
\usepackage{bbold}

\begin{document}
\title{Heart and diamond Fermi arcs in Pd and Pt oxide topological Dirac semimetals}
\author{Gang Li}
\email{gangli.phy@gmail.com}
\affiliation{Institute of Solid State Physics, Vienna University of Technology, A-1040 Vienna, Austria}
\author{Binghai Yan}
\affiliation{Max Planck Institute for Chemical Physics of Solids, 01187 Dresden, Germany}
\affiliation{Max Planck Institute for the Physics of Complex Systems, 01187 Dresden, Germany}
\affiliation{School of Physical Science and Technology, ShanghaiTech University, Shanghai 200031, China}
\author{Karsten Held}
\affiliation{Institute of Solid State Physics, Vienna University of Technology, A-1040 Vienna, Austria}                                       

\begin{abstract}
\textbf{
Topological Dirac semimetals (DSMs) exhibit nodal points through which energy bands disperse linearly in three-dimensional (3D) momentum space, a 3D analogue of graphene. The first experimentally confirmed DSMs with a pair of Dirac points (DPs), Na$_{3}$Bi and Cd$_{3}$As$_{2}$, show topological surface Fermi arc states and exotic magneto-transport properties, boosting the interest in the search for stable and nontoxic DSM materials. Here, based on the {\it ab-initio} band structure calculations we predict a family of palladium and platinum oxides as robust 3D DSMs. 
The novel topological properties of these compounds are distinct from all other previously reported DSMs,  they are rendered by the multiple number of DPs in the compounds. We show that along three unit axes of the cubic lattice there exist three pairs of DPs, 
which are linked by Fermi arcs on the surface. 
The Fermi arcs display a Lifshitz transition from a heart- to a diamond-shape upon varying the chemical potential. Corresponding oxides are already available as high-quality single crystals, which is an excellent precondition for the verification of our prediction by photoemission and magneto-transport experiments. 
Our findings not only predict a number of much robust 3D DSM candidates but also extend DSMs to transition metal oxides, a versatile family of materials, which opens the door to investigate the interplay between DSMs and electronic correlations that is not possible in the previously discovered DSM materials.}
\end{abstract}

\maketitle
\section*{Introduction}
The Dirac semimetal (DSM) is a new quantum state of matter whose low-energy excitations can be described by massless Dirac fermions with a vanishing density of states at the Fermi level.
The prototype of a DSM in two dimensions (2D) is graphene~\cite{RevModPhys.81.109}, whose conduction and valence bands linearly cross at a node, called the Dirac point (DP). 
Electrons and holes are massless at the DPs, which can be effectively viewed as 2D Dirac fermions at low energies.
With the mushrooming development of topological insulators (TIs)~\cite{RevModPhys.82.3045, RevModPhys.83.1057}, 1D and 2D Dirac fermions are also realized as topological objects at the edge or surface of TIs, respectively.
However, the search for 3D DSMs in bulk materials encounters strong difficulties, especially in the presence of strong spin-orbital coupling (SOC). 
Owing to the fact that the conduction and valence bands are spin degenerate, they can hybridize and gap out at the DPs. 
Only in the past few years, it was learned that the 3D DPs can be stabilized when these points are protected by additional symmetry that forbids the interband coupling~\cite{Yang:2014fk}, for which one example is the crystalline symmetry. 
The 3D DSM phase discovered in both Na$_{3}$Bi~\cite{PhysRevB.85.195320, Liu21022014} 
and Cd$_{3}$As$_{2}$~\cite{PhysRevB.88.125427,PhysRevLett.113.027603,Ncomms4786,Liu2014Cd3As2} and with one pair of DPs belongs to this class. 

The Dirac point can split into two Weyl nodes when the band degeneracy is lifted by breaking either time-reversal symmetry (TRS) or inversion symmetry (IS), leading to the Weyl semimetal (WSM)~\cite{Wan2011}. 
These Weyl points always come in pairs which carry opposite chiralities (handedness). 
One of the hallmark of WSMs is the appearance of the Fermi arc that connects a pair of two Weyl points with opposite chiralities, as found by recent theoretical~\cite{PhysRevX.5.011029,Huang2015,Sun2015arc} and experimental~\cite{PhysRevX.5.031013, Xu07082015, Yang2015TaAs,ZKLiu2016}  works on the TaAs-family of WSMs.
Therefore, the DSM can exhibit double Fermi arcs due to the degeneracy of Weyl points, which has been experimentally observed in Na$_{3}$Bi~\cite{Liu21022014,Xu16012015}. In addition, WSMs and DSMs usually show unconventional magneto-transport properties~\cite{Liang:2014ev,Narayanan:2015ed,Shekhar2015,Huang2015anomaly,Zhang2015ABJ,Xiong:2015kl}, like the chiral anomaly effect~\cite{Adler1969,Bell1969}.

So far all the experimentally discovered 3D DSMs contain only a single pair of DPs, which exhibits only one specific type of  Fermi arcs. 
Important questions arise for instances whether different forms of Fermi arcs in 3D DSMs can coexist and how they interact with each other~\cite{PhysRevB.90.205136,2015arXiv150902180K,2015arXiv151201552F}. To this end, a system with multiple pairs of DPs is urgently needed.
Here, we propose a family of palladium and platinum oxides that possesses three pairs of DPs which demonstrates different types of Fermi arcs in one single system.  
Many of these compounds are already available in the single crystal form which provides an ideal starting point for the systematic study of the novel electronic, thermo and magnetic properties of Fermi arcs and their interactions. 
Our work also significantly extends the topological concept into the versatile family of transition-metal oxides, which may serve as a robust material playground for the interplay between the 3D DSM and electronic correlations~\cite{PhysRevB.90.075137, PhysRevB.92.235149}.  

\section*{Electronic structure and topological nature}
\begin{figure}[htbp]
\centering
\includegraphics[width=\linewidth]{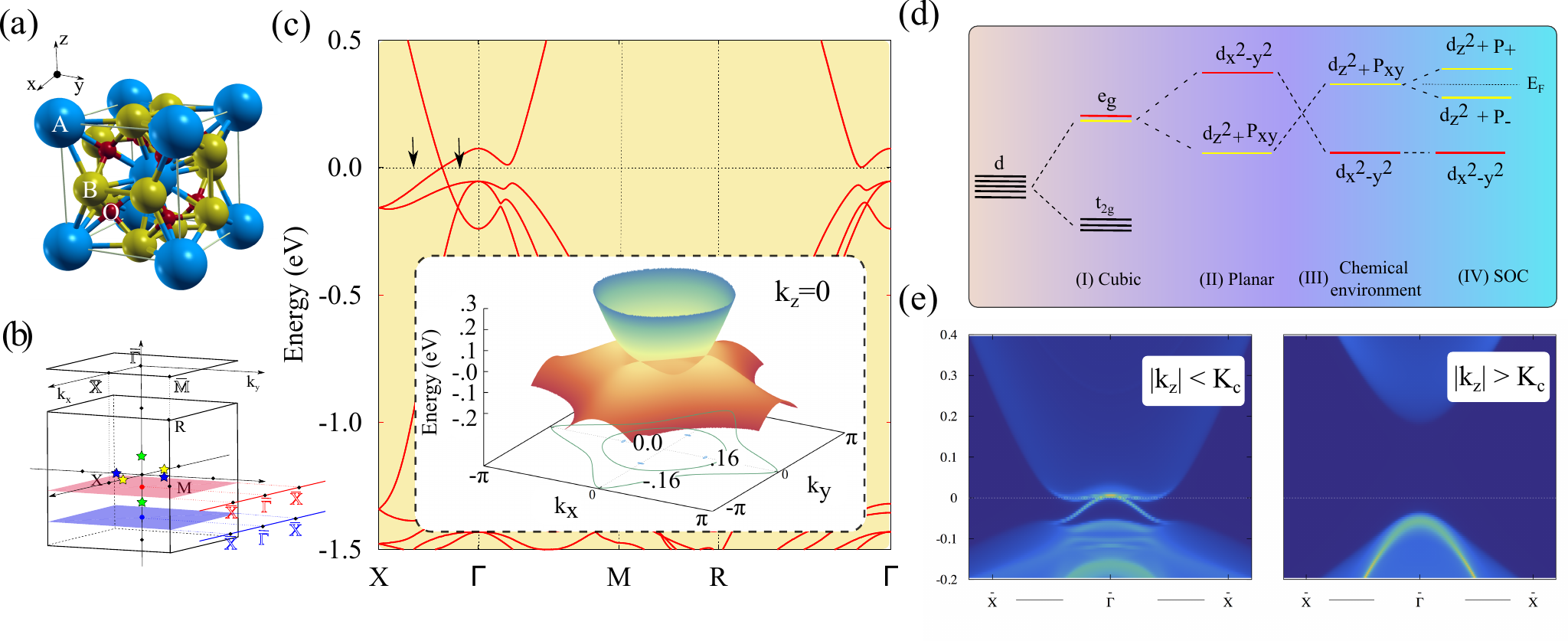}
\caption{\textbf{Electronic structure of AB$_{3}$O$_{4}$}. 
\textbf{(a)} Crystal structure of AB$_{3}$O$_{4}$, which contains A atoms (blue) at the corner, B atoms (beige) on each surface which are surrounded by four coplanar oxygen ions (red). 
\textbf{(b)} First Brillouin Zone with the three pairs of DPs (coloured stars). The pink and light blue plane are 2D cuts of the 3D BZ with $|k_{z}| < k_{z}^{c}$ (pink) and $|k_{z}| > k_{z}^{c}$ (light blue).
These two $k_{z}$ values are roughly indicated by arrows in Fig.~\ref{TDS}(\textbf{c}).  
\textbf{(c)} Electronic structure of prototypical SrPd$_{3}$O$_{4}$. 
Between X and $\Gamma$ the conduction and valence bands linearly cross at the Fermi level, featuring a 3D DP. There are in total three pairs of DPs located at the $k_{x}$, $k_{y}$ and $k_{z}$ axes. The inset shows the highest valence band and the lowest conduction bands which touch at four DPs in the $k_{z}=0$ plane. \textbf{(d)} The DPs are induced by a band inversion at $\Gamma$ formed by Pd (Pt) $e_{g}$-orbital and its hybridisation with the $p$-orbital of the oxygen atoms. \textbf{(e)} The topological nature is characterised by different $Z_{2}$ invariants at the two 2D planes shown in \textbf{(b)}. Hence topological edge states are found at the edge of a plane with $|k_{z}|< k_{z}^{c}$, while they are missing at the edge of a plane with $|k_{z}|>k_{z}^{c}$.}
\label{TDS}
\end{figure}

Here we theoretically investigate one of the most widely studied phase of noble metal oxides AB$_{3}$O$_{4}$ (A = Li, K, Na, Mg, Ca, Sr, Ba, Zn and Cd; B = Pd and Pt).
They crystallize in the space group Pm$\overline{3}$n (No. 223) and have a cubic NaPt$_{3}$O$_{4}$-type crystal structure, as shown in Fig.~\ref{TDS}\textbf{(a)}. 
These palladium and platinum atoms at the corner of the cubic cell are located between four coplanar oxygen ions, forming corner-sharing PdO$_{4}$ and PtO$_{4}$ tetragons.   
(The lattice parameters for all compounds that have been reported in the literatures can be found in Table~\ref{Table}). 
Here, we will show, through first-principle calculations, that these systems are 3D DSMs with three pairs of DPs that leads to the coexistence of different forms of Fermi arcs in one system.

\begin{table*}[htbp]
\centering
\begin{tabular}{|c|c|c|c|c|}
\hline
A &Pd ($a$, \AA) & Pd (DP, meV) & Pt ($a$, \AA) & Pt (DP, meV)\cr
\hline
Li & $\times$ & $\times$ & 5.634 ~\cite{MaterialRef} & 227 \cr
\hline
K & $\times$ & $\times$ & 5.679 ~\cite{MaterialRef}  &  225 \cr
\hline
Na & 5.655~\cite{MaterialRef}, 5.650~\cite{MaterialRef2} & 202  & 5.670~\cite{MaterialRef, MaterialRef3}, 5.687~\cite{MaterialRef4} & 281 \cr
\hline
Mg & $\times$ & $\times$ & 5.640~\cite{MaterialRef} & 10 \cr
\hline
Ca & 5.747~\cite{MaterialRef}, 5.747~\cite{doi:10.1021/ba-1971-0098.ch003} & 44  & 5.743~\cite{MaterialRef} & 45 \cr
\hline
Sr & 5.825~\cite{MaterialRef}, 5.822~\cite{MaterialRef5}, 5.826~\cite{doi:10.1021/ba-1971-0098.ch003} & 0.0  & $\times$ & $\times$ \cr 
\hline
Ba & $\times$ & $\times$ & 5.715~\cite{MaterialRef} & 11  \cr
\hline
Zn & $\times$ & $\times$ & 5.649~\cite{MaterialRef} & 52 \cr
\hline
Cd & 5.740~\cite{MaterialRef, CaPd3O4Ref}, 5.742~\cite{doi:10.1021/ba-1971-0098.ch003} & 0.0  & 5.687~\cite{MaterialRef} & 0.0  \cr
\hline
\end{tabular}
\caption{\textbf{Lattice parameters $a$ (in \AA) and DP position (in meV) relative to the Fermi level for various palladium and platinum oxides.} }
\label{Table}
\end{table*}

We found that the DPs in SrPd$_{3}$O$_{4}$, CdPd$_{3}$O$_{4}$ and CdPt$_{3}$O$_{4}$ are right at the Fermi level, in the following we choose SrPd$_{3}$O$_{4}$ as a representative for our discussion. 
The DPs in the other systems are slightly above the Fermi level, see Tab.~\ref{Table}. 
We first show that these systems host three pairs of DPs.
Fig.~\ref{TDS}\textbf{(c)} displays the electronic structure of SrPd$_{3}$O$_{4}$ with spin-orbital coupling (SOC), for the corresponding 3D Brillouin zone (BZ) see Fig.~\ref{TDS}\textbf{(b)}.  
As it can be clearly seen, the overall electronic structure of SrPd$_{3}$O$_{4}$ features itself as a 3D DSM with a linear band crossing, {\it i.e.}, a DP, at the Fermi level between X and $\Gamma$.  
In contrast to hitherto discussed 3D DSMs, there are six equivalent X points; the DPs between X and $\Gamma$ form three pairs located at $k_{x}, k_{y}$ and $k_{z}$ axes, respectively. 
In the inset of Fig.~\ref{TDS}\textbf{(c)} a 3D plot of the highest valence and the lowest conduction bands for $k_{z}=0$ plane is shown, where the two pairs of DPs at the $k_{x}$ and $k_{y}$ axes can be clearly seen. 
 
Next, we demonstrate that the DPs in these systems are symmetry protected, hence they are stable DPs.
In addition to the TRS and IS, a 4-fold rotational symmetry followed by a half-lattice translation along the screw axis protects these DPs. 
AB$_{3}$O$_{4}$ system with space group Pm$\overline{3}$n preserves TRS and IS,
which leads to their doubly degenerated electronic structures at each $\vec{k}$.
The Pd-Pd (Pt-Pt) chain at the (010) surface is a screw axis, which is perpendicular to the $xy$-plane. 
Under the screw rotation operation, the Hamiltonian transforms as $S_{4}^{z}H(k_{x}, k_{y}, k_{z}){S_{4}^{z}}^{-1}=H(-k_{y}, k_{x}, -k_{z})$, where $S_{4}^{z}$ is the screw rotation symmetry with the 4-fold rotation about Pd-Pd (Pt-Pt) chain along the $z$-axis followed by a half-lattice translation in the $z$-direction.    
Under the joint operations of TRS and $S_{4}^{z}$, the Hamiltonian is invariant as a function of $k_{z}$, {\it i.e.}, $(TS_{4}^{z})H(k_{x}, k_{y}, k_{z})(TS_{4}^{z})^{-1}=H(k_{y}, -k_{x}, k_{z})$. 
In this case, we can express the Hamiltonian for $k_{x}=k_{y}=0$ as $H(k_{z})|_{k_{x}=k_{y}=0}=d_{0} + d_{1}\sigma_{3} + d_{2}\tau_{3}\sigma_{3} + d_{3}\tau_{3}$ with $\sigma_{i}$ and $\tau_{i}$ being the Pauli matrices in spin and orbital spaces, $d_{i}$ are real coefficients~\cite{PhysRevLett.108.140405,Yang:2014fk, 2015arXiv150707504G,PhysRevB.90.205136}.  
With the constraint of double degeneracy and opposite spin direction on each state, only the $d_{3}(k_{z})=M_{0}-M_{1}k_{z}^{2}$ ($M_{0}$ and $M_{1}$ are real coefficients) term in the Hamiltonian survives, which leads to the two DPs at $k_{z}^{c}=\pm\sqrt{M_{0}/M_{1}}$ with $H(0,0,k_{z}^{c})=0$. 
A similar analysis can be carried out for $S_{4}^{x/y}$ with TRS and IS, which will give rise to the other two pairs of DPs at the $k_{x}$ and $k_{y}$ axes.  
Due to the protection from the three screw axes symmetries, the DPs in palladium (platinum) oxides AB$_{3}$O$_{4}$ are stable. 
Only by breaking the screw axes symmetry, they can be gaped out.   
For example, by applying compressive strain along $z$-direction, the screw axes symmetry at $x$ and $y$ direction is lost. 
An energy gap at the $k_{xy}$ plane is opened (results are not shown here), leaving the system with only a single pair of DPs at the $k_{z}$ axis. 
As will be discussed in the next paragraph, there is a band inversion in these systems which is unaffected by these screw axes operations. 
It is thus possible to drive these 3D DSMs to strong TIs by completely gaping out the DPs, which can be achieved for instance by applying different strains along $x$, $y$ and $z$ directions. 

Furthermore, we show that the DSM phase discovered in AB$_{3}$O$_{4}$ is topologically nontrivial. 
The topology of the DSM phase can be understood from the inverted band structure at $\Gamma$. 
To establish the topology, we show the energy splitting of the bands at $\Gamma$ in Fig.~\ref{TDS}\textbf{(d)}.
In AB$_{3}$O$_{4}$, Pd and Pt are in their $d^{8}$ valence configuration which is split into $t_{2g}$ and $e_{g}$ multiplets by the large cubic crystal field. 
The $t_{2g}$ states are completely filled while the $e_{g}$ states are half occupied.
The coplanar field formed by Pd (Pt) and the surrounding oxygen atoms further splits the $e_{g}$ states into $d_{x^{2}-y^{2}}$ and $d_{z^{2}}$ with $d_{x^{2}-y^{2}}$ being above $d_{z^{2}}$ in their normal order.
However, in AB$_{3}$O$_{4}$ the order of the two states is inverted at $\Gamma$, {\it i.e.} $d_{x^{2}-y^{2}}$ is below $d_{z^{2}}$ as shown in Fig.~\ref{TDS}\textbf{(d III)}.
Without SOC, the $d_{z^{2}}$ state stays at the Fermi level and $d_{x^{2}-y^{2}}$ would be fully occupied.   
As it is well known the effect of SOC on $e_{g}$ states is quenched.
While with SOC, Pd (Pt) $d_{z^{2}}$ states can hybridise with O $p_{x,y}$ states, which eventually opens the gap at $\Gamma$. 

In the following we will further show that, the inverted band structure also leads to another marvellous phenomenon in these topological DSMs -- the appearance of topological edge states (TES) at the edge of a 2D cut of the 3D BZ.
This serves as another independent proof for the topological nature of these systems. 
In Fig.~\ref{TDS}\textbf{(b)} two 2D cuts at different $k_{z}$ are shown as examples. 
When $|k_{z}|< k_{z}^{c}$ the $d_{x^{2}-y^{2}}$ and $d_{z^{2}}$ bands are inverted at $\Gamma$ while when $|k_{z}|> k_{z}^{c}$ they are not.
Thus, the two different 2D cuts carry different 2D topological indices. 
Here we show that such 2D topological index is exactly the 2D $Z_{2}$ topological invariant defined in 2D TIs protected by TRS. 
The palladium and platinum oxides AB$_{3}$O$_{4}$ preserves TRS and mirror symmetry. Under their joint operation the Hamiltonian is invariant for fixed $k_{z}$, {\it i.e.}, $(TM_{z})H(k_{x}, k_{y}, k_{z})(TM_{z})^{-1}=H(-k_{x}, -k_{y}, k_{z})$. 
Thus, $TM_{z}$ effectively serves as the TRS for the 2D cuts with fixed $k_{z}$ in the 3D BZ~\cite{PhysRevB.89.235127}.
This validates the definition of the 2D $Z_{2}$ topological invariant for the coloured planes shown in Fig.~\ref{TDS}\textbf{(b)}. 
2D systems preserving TRS host TES when $Z_{2}=1$. 
Thus, the TES are supposed to appear along the edges of planes with $|k_{z}| < k_{z}^{c}$.  
We calculate the TES by using the recursive Green's function approach~\cite{0305-4608-15-4-009} for the edge of the pink and the light blue planes shown in Fig.~\ref{TDS}\textbf{(b)}. 
The corresponding TES results are shown in Fig.~\ref{TDS}\textbf{(e)}. 
For a plane with $|k_{z}|<k_{z}^{c}$ the TES connecting the valence and conduction bands are clearly visible, while for $|k_{z}|> k_{z}^{c}$ the system is a trivial insulator with band a gap.
The TES again justify the topological nature of these 3D DSMs, consistent with the above band inversion analysis.
The same topological behaviour holds as well for the other two pairs of DPs at the $k_{x}$ and $k_{y}$ axes, because they are symmetrically related.

\section*{Fermi arcs and Lifshitz transition}
Topological DSMs host nontrivial open Fermi surfaces connecting opposite chiral charges.  
In the Pd and Pt oxides proposed in this work, we discover multiple pairs of DPs. 
Each DP can be viewed as two copies of Weyl nodes with opposite chiral charges. 
Fermi arcs may appear and connect one positive Weyl node with one negative Weyl node. 
In contrast to previous works, we have however more than one pair of DPs.
In principle, any positive Weyl node can be connected to any other negative Weyl node in the system. 
Therefore, the question of how the opposite chiral charges among these DPs are connected by Fermi arcs naturally arises. 
The Fermi arcs in DSMs are protected by symmetry, they can be different on different surfaces depending on the surface symmetry and potential. Here, we choose the (001) surface to answer this question and demonstrate that two different types of Fermi arcs coexist in these systems, which has not been reported before in the literature.  
 
\begin{figure}[htbp]
\centering
\includegraphics[width=\linewidth]{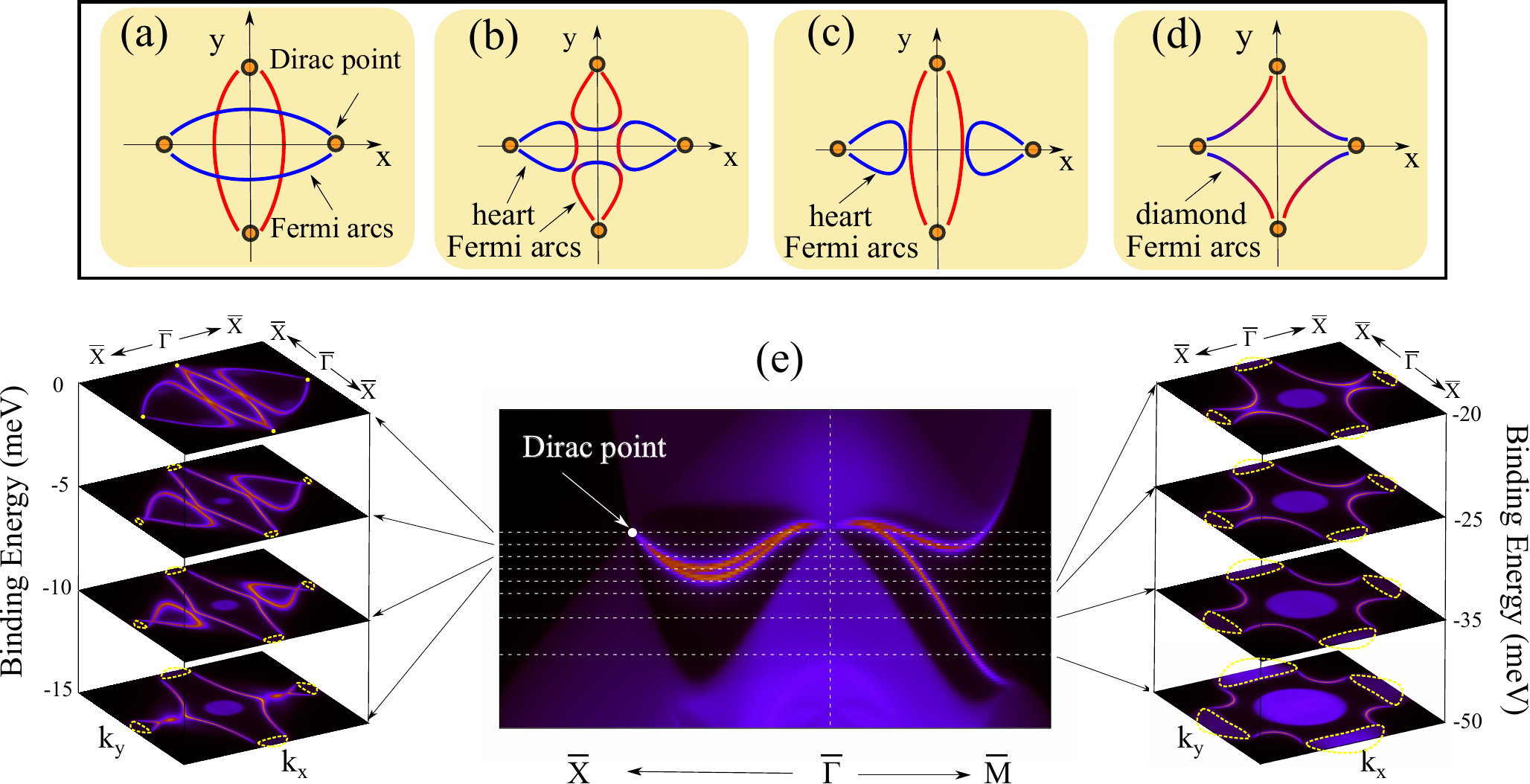}
\caption{\textbf{``Double Fermi arcs" of noble metal oxides AB$_{3}$O$_{4}$.} \textbf{(a)-(d)} Four symmetry-allowed Fermi arcs on the (001) surface connect the opposite chiral charges at the DPs, which can display both heart and diamond shapes. \textbf{(e)} The topological surface states on (001) surface are shown along two high-symmetry paths (see Fig.~\ref{TDS}\textbf{(b)} for the BZ). Below the Fermi level, eight different constant energy cuts are taken to illustrate the Fermi arcs in the entire surface BZ and their evolution. The position of the bulk DPs are indicated by yellow ellipse.}
\label{FermiArcs}
\end{figure}

Before discussing the detailed Fermi arc pattern, we would like to provide a simple argument on the possible constraints. 
Generally, a surface does not necessarily preserves the symmetries that protect the bulk DPs. 
However for the Pd and Pt oxides AB$_{3}$O$4$, mirror symmetry is also preserved on the (001) surface. 
Let us additionally consider $C_{2}$ rotational symmetry, which is preserved on the surface as well and imposes additional constraint on the Fermi arcs. 
The top surface (001) is parallel to the $\mathbf{a}$ and $\mathbf{b}$ lattice vectors.
It remains unchanged under the $C_{2}$ rotational symmetry and mirror symmetry, {\it i.e.}, $(x, y, z)\xrightarrow{C_{2}} (-x, -y, z)$, $(x, y, z)\xrightarrow{M_{x}}(-x, y, z)$, $(x, y, z)\xrightarrow{M_{y}}(x, -y, z)$.
In momentum space, the two DPs along $k_{z}$ axis project to the same surface $\overline{\Gamma}$ point, only the other two pairs of DPs located at the $k_{x}$ and $k_{y}$ axes remain separated on the (001) surface. 
The Hamiltonian of the Fermi arcs $H_{arc}$ for these two pairs of DPs transforms under these two symmetries as $C_{2}H_{arc}(k_{x}, k_{y}, k_{z})C_{2}^{-1}=H_{arc}(-k_{x}, -k_{y}, k_{z})$,  
$M_{x}H_{arc}(k_{x}, k_{y}, k_{z})M_{x}^{-1}=H(-k_{x}, k_{y}, k_{z})$ and $M_{y}H_{arc}(k_{x},k_{y}, k_{z})M_{y}^{-1}=H_{arc}(k_{x}, -k_{y}, k_{z})$.
Together with the necessity that a Weyl node with negative chirality connects to a Weyl node with positive chirality, this allows us to identify the four possible types of Fermi arcs illustrated in Fig.~\ref{FermiArcs}\textbf{(a)-(d)}.
Here, the Fermi arcs with the same colour transform onto each other under $C_{2}$ and mirror symmetry. 
In Fig.~\ref{FermiArcs}\textbf{(a)}, the two DPs along the same axis are connected by two Fermi arcs. These are the ordinary Fermi arcs observed also in Na$_{3}$Bi~\cite{Liu21022014,Xu16012015}. 
But since we have more than one pair of DPs, the intersection of two ordinary Fermi arcs can hybridise and deform into the other types, such as the heart and diamond Fermi arcs shown in Fig.~\ref{FermiArcs}\textbf{(b)-(d)}. 

This is confirmed by our direct calculation of the surface states shown in Fig.~\ref{FermiArcs}\textbf{(e)}, which implies that the Pd and Pt oxides AB$_{3}$O$_{4}$ host the Fermi arcs shown in Fig.~\ref{FermiArcs}\textbf{(c)} and \textbf{(d)}.
In the middle panel of Fig.~\ref{FermiArcs}(\textbf{e}) we plot the topological surface states along two high-symmetry paths $\overline{\Gamma}\rightarrow\overline{X}$ and $\overline{\Gamma}\rightarrow\overline{M}$. 
The bands with stronger intensity (red) are at the surface, while the shadow bands in the background are of bulk character. 
The bulk valence and conduction bands touch at the nodal point between $\overline{\Gamma}$ and $\overline{X}$, featuring the 3D DPs at the $k_{x}$ and $k_{y}$ axes. 
As is clearly shown, there are topological surface states connecting the bulk valence and conduction bands and they behave differently along the two chosen high-symmetry paths. 

Let us now change the chemical potential to further demonstrate that they correspond to two different types of Fermi arcs in these systems, which have not been observed in previously proposed DSMs. 
To this end, on the left and the right hand sides of Fig.~\ref{FermiArcs}(\textbf{e}), plots of Fermi arcs in the entire 2D surface BZ at eight chosen binding energies are shown. 
As the axis of the bulk Dirac cone is tilted, its intersection with constant energy plane displays an ellipse shape, which is indicated in yellow in these plots.   
At the Fermi level, Fermi arcs display a Fig.~\ref{FermiArcs}\textbf{(c)}-type pattern.
The two DPs at the $k_{y}$ axis are connected by open Fermi surfaces. 
This corresponds to the same type of Fermi arcs observed already in Na$_{3}$Bi.
On top of this we observe, however, another heart-shape Fermi surface around the DPs at the $k_{x}$ axis connecting the two opposite chiral charges from the same DP.
These two heart-shape Fermi surfaces are symmetric with respect to the surface $\overline{\Gamma}$ point and $k_{y}$ axis. 
Note that both branches of the heart-shape Fermi surfaces are disconnected at the DPs at $k_{x}^{c}$ and $-k_{x}^{c}$. 
Thus, these are also open Fermi arcs. 
Let us emphasise that these are different Fermi arcs as compared to the ones parallel to the $k_{y}$ axis.
They are iso-energy projections of the Riemann surface states recently proposed in Ref.~\cite{2015arXiv151201552F}, which we denote as ``Riemann Fermi arcs".
As described by holomorphic functions, the Riemann surface states manifest themselves by covering the complex plane a finite or infinite number of times.  
As a result, the intersection of a helicoid and an anti-helicoid Riemann surface at the Dirac cone discriminates the ``Riemann Fermi arcs" from the ordinary Fermi arcs, as the latter stay always separated in the surface BZ. 
Here we demonstrate their difference in an energy evolution plot of the Fermi surface at -5 meV, -10 meV and -15 meV, respectively.
Continuously decreasing the energy significantly modifies the heart-shape Fermi surfaces connecting the opposite chiral charges at each DP at the $k_{x}$ axis, while the Fermi arcs connecting the two DPs at the $k_{y}$ axis remains topological unchanged.
This is clearly demonstrates the difference between the two types of Fermi arcs. 

We note that the shrinking of the heart-shape Fermi arcs results from the intersection of the helicoid and the anti-helicoid Riemann surfaces along the $k_{x}$ axis at the Dirac cone. 
The intersection of them always appears at the $k_{x}$ axis, which is required by the $C_{2}$ rotational symmetry and mirror symmetry preserved on this surface. 
Unlike WSMs, the Fermi arcs in 3D DSMs are not stable against symmetry-allowed perturbations.
They can be continuously deformed into other forms without breaking any symmetry~\cite{PhysRevB.90.205136,2015arXiv150902180K,2015arXiv151201552F}.  
By further decreasing energy, we indeed observe a Lifshitz transition of the Fermi surface, which is shown on the right hand side of Fig.~\ref{TDS}\textbf{(e)}.   
Here, two types of Fermi arcs now merge into one single type of Fermi arcs shown in Fig.~\ref{TDS}\textbf{(d)}, {\it e.g.}, the opposite chiral charges locating at $k_{x}^{c}$ and $k_{y}^{c}$ are now connected by one open Fermi surface. 
Under $C_{2}$ rotation and mirror symmetries, alltogether four copies of it show up in the entire 2D surface BZ forming a diamond shape structure.  
With the further decrease of energy, the length of this type of Fermi arcs becomes smaller, since the Dirac cone gradually expands. 

\section*{Experimental indications}
In addition to the above theoretical investigations, let us here reexamine the experimental resistivity and thermopower data to show that they may provide already a first experimental support for SrPd$_{3}$O$_{4}$ and CaPd$_{3}$O$_{4}$ being topological nontrivial.
It is known that ternary Pd and Pt oxides have been extensively studied many years ago owing to a completely different reason, {\it i.e.}, as fuel-cell electrocatalysts and for thermoelectric applications~\cite{doi:10.1021/ja016522b, 2015ZNatA..70..240S}. 
SrPd$_{3}$O$_{4}$ and CaPd$_{3}$O$_{4}$ single crystals have been synthesized in high quality~\cite{Smallwood2000299,CaPd3O4Ref, MaterialRef5, Taniguchi200467, PhysRevB.68.233101,Itoh1999715, doi:10.1021/ba-1971-0098.ch003}.
Experimental results on both thermo- and electronic-transport are available. 
However, the previous experimental understanding of the resistivity and thermopower in SrPd$_{3}$O$_{4}$ and CaPd$_{3}$O$_{4}$ seem to be confusing and contradictory to each other from a conventional viewpoint.
As a consequence, these systems were concluded reluctantly to be trivial semiconductors with a small energy gap.  
Here, we readdress this old issue from the topological perspective established in this work and we show that these puzzles can now be understood consistently within the topological picture.  

\begin{figure}[htbp]
\centering
\includegraphics[width=\linewidth]{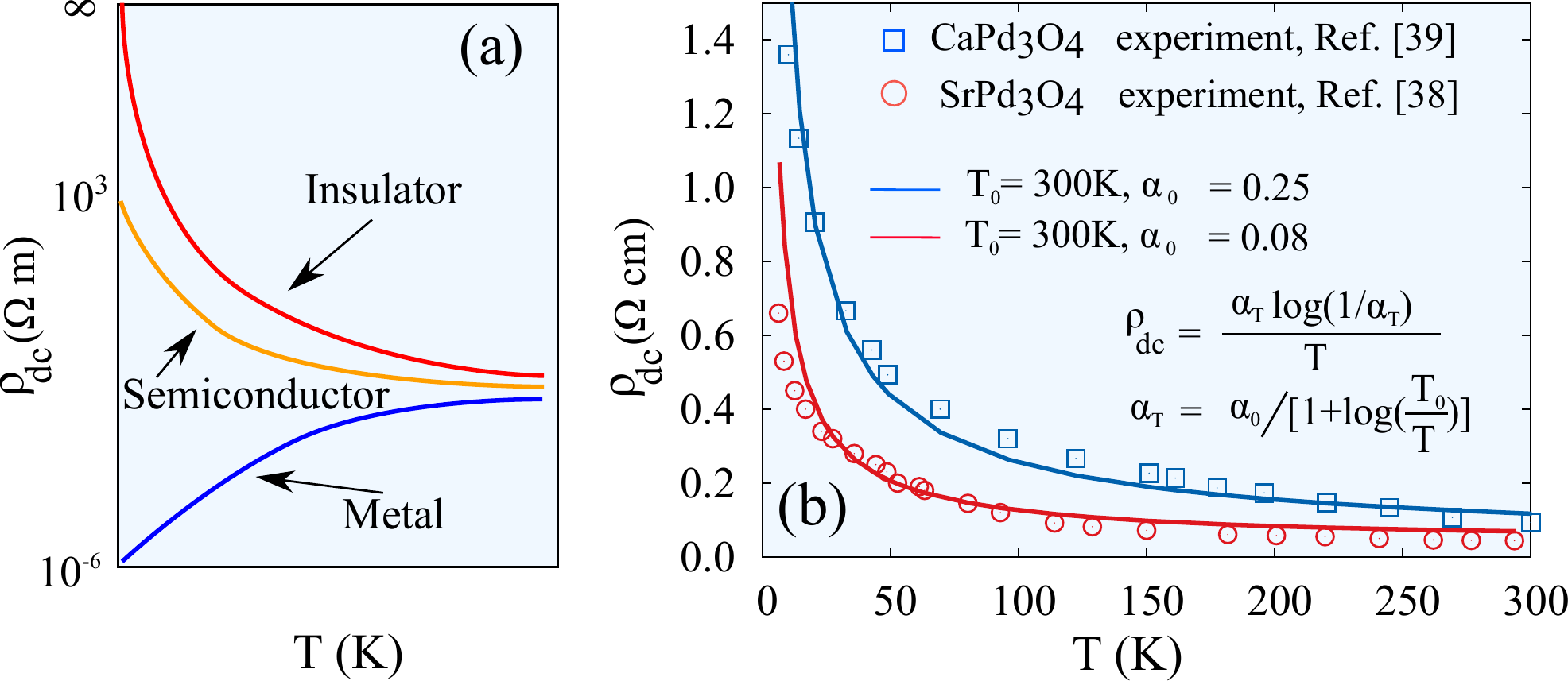}
\caption{\textbf{Resistivity anomaly in correlated semimetals.} \textbf{(a)} Schematic plot of the  typical behaviors of resistivity in an insulator, semiconductor and metal. \textbf{(b)} Small residual resistivity at $T\rightarrow0$ and the temperature dependence of the resistivity in SrPd$_{3}$O$_{4}$ and CaPd$_{3}$O$_{4}$ can be explained by a DSM with long-range Coulomb interaction, see Refs.~\cite{PhysRevLett.107.196803, PhysRevLett.108.046602}.}
\label{resistivity}
\end{figure}

\textit{Electric resistivity anomaly:} The electric resistivity, a measure of the charge carrier density, is one of the most intuitive transport quantities that distinguishes metals from insulators. 
In metals, the resistivity takes value less than a few $\mu\Omega\cdot m$ and it increases with increasing temperature. 
In insulators, at absolute zero-temperature there are no free conduction electrons, the resistivity becomes infinite. 
Semiconductors, which can be viewed as an insulator with small energy gap, contains a few charge carriers so that they have a large but finite resistivity ($\sim10^{3} \Omega\cdot m$). 
 A schematic plot of the typical behaviors of resistivity can be found in Fig.~\ref{resistivity}\textbf{(a)}.
The experimentally confirmed DSMs, {\it i.e.}, Cd$_{3}$As$_{2}$ and Na$_{3}$Bi, display a clear metallic behavior with a tiny resistivity at zero temprature~\cite{Liang:2014ev, Kushwaha:2015}, indicating an extremely high mobility of charge carriers in these systems. 
For the DSMs proposed in this work, a similar metallicity is hence expected. 
Surprisingly, the resistivity of SrPd$_{3}$O$_{4}$ and CaPd$_{3}$O$_{4}$ is finite and decreases with increasing temperatures, e.g., see Fig.~\ref{resistivity}\textbf{(b)}.
At first glance, this seems to support them to be trivial semiconductors.
However, it is crucial to note that the resistivity for $T\rightarrow0$ is four to five orders smaller than typical values for semiconductors, it is only $\sim10^{-3}-10^{-2}\Omega\cdot m$ in SrPd$_{3}$O$_{4}$~\cite{MaterialRef5} and CaPd$_{3}$O$_{4}$~\cite{CaPd3O4Ref, PhysRevB.68.233101}.

We note that the small residual resistivity and the low-temperature behaviour of the resistivity in SrPd$_{3}$O$_{4}$ and CaPd$_{3}$O$_{4}$ may imply that they are semimetals with strong electronic correlations. 
To this end, we want to recall the similar resistivity anomaly in the topological Weyl SM Y$_{2}$Ir$_{2}$O$_{7}$~\cite{doi:10.1143/JPSJ.70.2880},
which is $\sim3\times10^{-2}\Omega\cdot m$ at zero temperature and also decreases with increasing temperature.
Both the residue and the temperature dependence of the resistivity in Y$_{2}$Ir$_{2}$O$_{7}$  highly resemble that of SrPd$_{3}$O$_{4}$~\cite{MaterialRef5} and CaPd$_{3}$O$_{4}$~\cite{CaPd3O4Ref, PhysRevB.68.233101}.
It is found that the Coulomb interaction can relax the conductivity of electrons in topological SMs, as a result a low temperature linear-T dependence in the DC conductivity was found~\cite{PhysRevLett.108.046602}.
This successfully explains the low-temperature resistivity anomaly of Y$_{2}$Ir$_{2}$O$_{7}$.
The same resistivity behaviour observed in SrPd$_{3}$O$_{4}$ and CaPd$_{3}$O$_{4}$ is very likely due to the same reason owing to the half-filled palladium $e_{g}$ orbitals.  
We thus, in close analogy to what has been done for the Weyl SM Y$_{2}$Ir$_{2}$O$_{7}$, fit the experimental data of SrPd$_{3}$O$_{4}$ and CaPd$_{3}$O$_{4}$ with a simplified resistivity expression for DSMs in the presence of long-range Coulomb interactions~\cite{PhysRevLett.107.196803}.
As shown in Fig.~\ref{resistivity}\textbf{(b)}, both the low- and high-temperature resistivity can be consistently reproduced. 
Based on the limited number of available 3D SMs, we summarise that the resistivity of SMs seem to follow essentially the metallic behaviour, but becomes finite at zero-temperature if the system is subject to long-range Coulomb interactions. 
In Cd$_{3}$As$_{2}$ and Na$_{3}$Bi, the electronic correlation is weak, the resistivity thus becomes negligibly small.  
Y$_{2}$Ir$_{2}$O$_{7}$, as well as SrPd$_{3}$O$_{4}$ and CaPd$_{3}$O$_{4}$, contain however heavy-fermion and transition-metal elements. 
As a result their resistivity shows a semiconducting like behaviour while the zero-temperature resistivity is much smaller than that in typical semiconductors. 

{\it Room-temperature thermopower anomaly:} 

\begin{table*}[htbp]
\centering
\begin{tabular}{|c|c|c|c|c|c|}
\hline
SrPd$_{3}$O$_{4}$ & CaPd$_{3}$O$_{4}$ & Cd$_{3}$As$_{2}$~\cite{Spitzer} & Cd$_{3}$As$_{2}$~\cite{0022-3727-7-1-321} & Cd$_{3}$As$_{2}$~\cite{Blom19691299} & Cd$_{3}$As$_{2}$~\cite{ChengZhang:17202}\cr
\hline
28 & 80 & $-57\sim-61$ & $-70\sim-82$ & -84 & -70 \cr
\hline
\end{tabular}
\caption{\textbf{Comparison of Seebeck coefficients $S (\mu V/K)$ of SrPd$_{3}$O$_{4}$ and CaPd$_{3}$O$_{4}$ with that of the DSM Cd$_{3}$As$_{2}$}. The different signs of $S$ indicate that the charge carriers in SrPd$_{3}$O$_{4}$ and CaPd$_{3}$O$_{4}$ are holes, while they are electrons in Cd$_{3}$As$_{2}$.}
\label{Tab2}
\end{table*}

Thermo-conductivity experiments support the metallicity of SrPd$_{3}$O$_{4}$ and CaPd$_{3}$O$_{4}$, {\it i.e.}, their Seebeck coefficient $S$ linearly decreases with decreasing temperature in the  low $T$ regime. 
However, $S$ at room temperature takes slightly larger values ($\sim 28 \mu V/K$ for SrPd$_{3}$O$_{4}$~\cite{MaterialRef5} and $\sim 80 \mu V/K$ for CaPd$_{3}$O$_{4}$~\cite{CaPd3O4Ref, PhysRevB.68.233101}) than those typical for metals ($\sim 10 \mu V/K$).
At the same time, these values are also much smaller than those typical for semiconductors ({\it e.g.}, $\sim 450 \mu V/K$ for Si). 
Similar to the resistivity anomaly, the striking features of the low- and room-temperature Seebeck coefficient place SrPd$_{3}$O$_{4}$ and CaPd$_{3}$O$_{4}$ somewhere between conventional metals and semiconductors. 
We believe that this previously puzzling result again implies SrPd$_{3}$O$_{4}$ and CaPd$_{3}$O$_{4}$ to be semimetals,
which gains further support when we compare the room-temperature Seebeck coefficients $S$ of SrPd$_{3}$O$_{4}$ and CaPd$_{3}$O$_{4}$ to those of the known DSM Cd$_{3}$As$_{2}$. 
As shown in Tab.~\ref{Tab2}, $S$ is of similar order in all three compounds.
In addition, in Cd$_{3}$As$_{2}$ the absolute value of the Seebeck coefficient $S$ linearly decreases to zero with the decreasing temperature, which is again the same as that for SrPd$_{3}$O$_{4}$ and CaPd$_{3}$O$_{4}$. 
The similar amplitude of room-temperature $S$ and its low-T temperature dependence are very likely to suggest that SrPd$_{3}$O$_{4}$ and CaPd$_{3}$O$_{4}$ are more like topological semimetals than as trivial semiconductors interpreted before. 
As the experimental data available hitherto at best provides for an indirect proof of a DSM, our calculations call for a more direct experimental confirmation, {\it e.g.}, by angle-resolved photoemission spectroscopy.

In conclusion, our systematic study on the electronic structure of palladium and paltinum oxides AB$_{3}$O$_{4}$ with space group Pm$\overline{3}$n establishes this family of materials to be the first transition metal oxide family that hosts multiple pairs of DPs in one system. 
The coexistence of different forms of Fermi arcs further discriminates them from other DSMs discovered so far. 
Owing to the availability of clean single crystals, the rich and striking features of this family of materials make it an ideal platform to experimentally study multiple DPs and the interaction of their open Fermi surfaces. 
Successful transforming topological phase to the transition-metal oxides further opens the door for studying the interplay between electronic correlation and nontrivial topology. 
Our theoretical investigation paves the way for a future experimental study of the electronic, transport and spectroscopic properties.

\section*{Methods}

{\it Ab-initio} calculations were carried out within the full-potential linearized augmented plane-wave method~\cite{FP-LAPW} implemented in Wien2k~\cite{Wien2k}.
We use $K_{max}\mbox{RMT}=9.0$ and a $10\times10\times10$ k-mesh for the ground-state calculations, where RMT represents the smallest muffin-tin radius and $K_{max}$ is the maximum size of reciprocal-lattice vectors. 
The modified Becke-Johnson exchange potential~\cite{PhysRevLett.102.226401}  is used in all calculations.
The surface electronic structures are further calculated using the maximally localized Wannier functions (MLWFs)~\cite{Mostofi2008685} employing the WIEN2WANNIER29~\cite{wien2wannier90} interface.


\section*{Acknowledgements}
We acknowledge financial support  by the European Research Council under the European Union's Seventh Framework Program (FP/2007-2013)/ERC through grant agreement n.\ 306447 and the Austrian Science Fund (FWF) through project ID I 1395-N26 as part of the DFG research unit FOR 1346 (GL, KH).
Calculations have been done in part on the Vienna Scientific Cluster (VSC).

\section*{Contributions}
G.L. conceived the project and performed the calculations.  All authors analysed the results and wrote the manuscript.

\section*{Competing financial interests}
The authors declare no competing financial interests.

\section*{Corresponding author}
Correspondence to: Gang Li

\end{document}